\documentstyle[prl,aps]{revtex}

\begin{document}
\author{Jian-Qi Shen$^1$\footnote{E-mail address: jqshen@coer.zju.edu.cn}, Zhi-Chao Ruan$^1$,
and Sailing He$^{1,2}$\footnote{E-mail address: sailing@kth.se}}
\address{$^1$ Centre for Optical
and Electromagnetic Research, Joint Research Centre of Photonics
of the Royal Institute of Technology (Sweden) and Zhejiang
University, Zhejiang University, Hangzhou Yuquan 310027, P. R.
China\\
$^2$ Laboratory of Photonics and Microwave Engineering,
Department of Microelectronics and Information Technology,\\
Royal Institute of Technology, Electrum 229, SE-164 40 Kista,
Sweden}
\date{\today}
\title{Influence of the signal light on the transient optical properties of a four-level EIT medium}
\maketitle

\begin{abstract}
General formulae for the transient evolution of the susceptibility
(absorption) induced by the quantum interference effect in a
four-level N-type EIT medium is presented. The influence of the
signal light on the transient susceptibility for the probe beam is
studied for two typical cases when the strength of the coupling
beam is much greater or less than that of the signal field. An
interesting {\it level reciprocity} relationship between these two
cases is found.
\\ \\


\end{abstract}

\section{Introduction}
Recently, many theoretical and experimental investigations have
shown that the control of phase coherence in a multilevel atomic
ensemble will give rise to many novel and striking quantum optical
phenomena in the wave propagation of near-resonant light. These
phenomena and effects include the atomic coherent population
trapping (CPT)\cite{Arimondo}, laser without
inversion\cite{Harris,Ima} and electromagnetically induced
transparency (EIT)\cite{Harris2}. The idea of the atomic CPT was
first suggested by Orriols {\it et al.} in 1976\cite{Orriols}, and
experimentally demonstrated by Gray {\it et al.} in
1978\cite{Gray} and by Alzetta {\it et al.} in 1979\cite{Alzetta}.
In both CPT and EIT, two laser beams are involved in the quantum
interference effect. In an EIT medium, if there is only one
propagating resonant laser beam, it will be absorbed; but no laser
beam will be absorbed when two appropriate laser beams propagate
through the same medium ({\it i.e.}, the opaque medium is turned
into a transparent one). In CPT the two fields interacting with
the atoms have nearly the same strength and the quantum
interference effect arises from both fields. In EIT, however, one
of the propagating laser beams is much weaker than the
other\cite{Harris2,Lukin}. Thus, the interference effect in EIT
can be said to be driven by the stronger one of the two laser
beams. This stronger beam is called the {\it coupling} beam, and
the weaker beam termed the {\it probe} beam. Historically, the
foundations of EIT were laid by Kocharovskaya and Khanin in
1988\cite{Kocharovskaya} and independently by Harris in
1989\cite{Harris}. The first experimental observation of EIT was
performed by Harris {\it et al.} in 1991\cite{Boller}. Besides the
CPT explanation (in which the concept of dark state or
non-coupling state is essential to the theoretical mechanism), EIT
can also be interpreted by using the points of view of the
interference between dressed states\cite{Marangos}, the multiple
routes to excitation (multi-pathway interference)
model\cite{Moseley}, and some other explanations such as the
quantum-field formulation, where one uses Feynman diagram to
represent the interfering process in EIT\cite{Cohen}. Due to its
unusual quantum coherent characteristics, the discovery of EIT has
so far led to many new peculiar effects and
phenomena\cite{Harris,Ima,Gheri,Fleischhauer,Agarwal,Harris1993,Eberly},
some of which are believed to be useful for the development of new
techniques in quantum optics\cite{Harris2}. More recently, the
physical EIT effects observed experimentally include the ultraslow
light pulse propagation\cite{Hau,Kash}, superluminal light
propagation\cite{Wang}, light storage in atomic
vapor\cite{Liu,Phillips}, and atomic ground state
cooling\cite{Morigi}.

EIT arises from the atomic phase coherence and quantum
interference in the atomic transition process. In 1995, Li {\it et
al.} investigated the transient properties induced by the quantum
interference effect \cite{Li}. These properties include the
absorption for the probe field, transient gain without population
inversion and enhancement of dispersion in a three-level EIT
atomic medium when the coupling laser is switched on\cite{Li}.
Recently, Greentree {\it et al.} studied the turn-on and turn-off
dynamics (including the resonant and off-resonant transient
behaviors) of EIT in a three-level
 medium \cite{Greentree}. Note that all the above investigations are
associated with the three-level EIT system. Recent evidences have
shown that the giant non-linearities ({\it e.g.}, the enhancement
of nonlinear absorption for the probe light) exist in a four-level
coherent atomic  medium \cite{Yamamoto,Yan}. The large nonlinear
optical susceptibilities in a four-level EIT medium have some
novel applications, {\it e.g.}, the realization of an absorptive
two-photon optical switch, in which a laser pulse controls the
absorption of another laser field. In this process, the EIT system
absorbs two photons, but not one photon\cite{Yamamoto}. Due to
their novel effects and potential applications, four-level EIT
media have attracted attention recently. For example, in 1998 Ling
{\it et al}. considered the EIT effect in a four-level N-type
Doppler broadened media\cite{Ling}; Harris and Yamamoto described
a four-level EIT atomic system that exhibits greatly enhanced
third-order susceptibility, but has vanishing linear
susceptibility\cite{Yamamoto}; Based on the suggestion of Harris
and Yamamoto\cite{Yamamoto}, Yan {\it et al.} reported in 2001 an
experimental demonstration of absorptive two-photon switch by
constructive quantum interference in  a four-level atomic system
(such as the cold $^{87}$Rb atoms) \cite{Yan}.

Little has been done in the literature on the investigation of the
properties of the transient evolution in a four-level EIT medium.
The consideration of the transient properties of EIT media is of
importance due to their potential applications such as the
absorptive optical switch\cite{Li}, in which the transmission of a
highly absorptive medium is controlled dynamically by an
additional signal (switching) light. In the present paper, we
study the transient optical properties and behaviors (including
the influence of the signal light on the probe light) in a
four-level N-type EIT  medium by using the semiclassical theory.
First, we propose a general treatment for the transient evolution
of the probability amplitudes of atomic levels in the four-level
EIT medium. Then we treat two typical cases in which the strength
of the coupling light is much greater or less than that of the
signal (switching) light. The influence of the signal light on the
probe light is considered, as well as the transient optical
behaviors of the four-level EIT atomic system. It is found that
there exists a reciprocity relationship (called {\it level
reciprocity} in the present paper) between these two cases . The
giant, resonantly enhanced nonlinearity is also discussed briefly
when the linear susceptibility vanishes in such a four-level EIT
medium.

\section{General treatment for the transient evolution in the four-level EIT media}

In this section, we use a semiclassical theory to derive some
general formulae for the transient behaviors of the probability
amplitudes of a four-level EIT medium when a signal field is
switched on. Consider a four-level atomic ensemble interacting
with three optical fields, namely, the coupling beam, the probe
beam and the signal field, whose Rabi frequencies are denoted by
$\Omega_{\rm c}$, $\Omega_{\rm p}$ and $\Omega_{24}$,
respectively. The configuration of the four-level system is
depicted in Fig. 1. In such a four-level atomic ensemble of ``N''
type, levels $|1\rangle$ and $|2\rangle$ are the ground states,
$|3\rangle$ and $|4\rangle$ the excited states. The probe,
coupling and signal fields couple the level pairs
$|1\rangle$-$|3\rangle$, $|2\rangle$-$|3\rangle$,
$|2\rangle$-$|4\rangle$, respectively. In the present paper, we
assume that these three fields are all in resonance with the
corresponding level transitions, {\it i.e.}, there is no frequency
detuning of $\Omega_{\rm c}$, $\Omega_{\rm p}$ or $\Omega_{24}$.

In the interaction picture, the Hamiltonian for such a four-level
N-type atomic ensemble has the following form (with $\hbar=1$ for
simplicity)\cite{Yamamoto}
\begin{equation}
H=-\frac{1}{2}\left(\Omega_{\rm p}|1\rangle\langle3|+\Omega_{\rm
c}|2\rangle\langle3|+\Omega_{24}|2\rangle\langle4|\right)+{\rm
h.c.},
\end{equation}
where ${\rm h.c.}$ represents the Hermitian conjugation. The
Hamiltonian associated with the level decay is assumed to take the
form $ H_{\Gamma}=i\hat{\Gamma}$ with $\hat{\Gamma}={\rm diag}[0,
-{\gamma_{21}}/{2}, -{\Gamma_{3}}/{2}, -{\Gamma_{4}}/{2}]$, where
$\Gamma_{3}$ and $\Gamma_{4}$ denote the spontaneous decay rates
of levels $|3\rangle$ and $|4\rangle$, respectively, and
$\gamma_{21}$ the dephasing rate (nonradiative decay rate) of
level $|2\rangle$. Here $\gamma_{21}, \Gamma_{3}$ and $
\Gamma_{4}$ are non-negative constants. Thus, in the interaction
picture the Schr\"{o}dinger equation governing the above
four-level atomic ensemble is
\begin{equation}
i\frac{\partial}{\partial t}|\psi(t)\rangle=H_{\rm tot
}|\psi(t)\rangle    \label{Sch}
\end{equation}
with the total Hamiltonian $H_{\rm tot}=H+H_{\Gamma}$. Let
$|\psi(t)\rangle=a_{1}(t)|1\rangle+a_{2}(t)|2\rangle+a_{3}(t)|3\rangle+a_{4}(t)|4\rangle$.
Then according to the Schr\"{o}dinger equation (\ref{Sch}), the
probability amplitudes $a_{i}(t)$ ($i=1,2,3,4$) satisfy the
following set of equations
\begin{equation}
\left\{
\begin{array}{ll}
&  \dot{a}_{1}=\frac{i}{2}\Omega_{\rm p}a_{3},                  \\
&   \dot{a}_{2}=\frac{i}{2}\left(\Omega_{\rm
c}a_{3}+\Omega_{24}a_{4}\right)-\frac{\gamma_{21}}{2}a_{2},  \\
&  \dot{a}_{3}=\frac{i}{2}\left(\Omega^{\ast}_{\rm
c}a_{2}+\Omega^{\ast}_{\rm p}a_{1}\right)-\frac{\Gamma_{3}}{2}a_{3},   \\
&
\dot{a}_{4}=\frac{i}{2}\Omega^{\ast}_{24}a_{2}-\frac{\Gamma_{4}}{2}a_{4},
\end{array}
\right. \label{eqamplitude}
\end{equation}
where dot denotes the derivative of $a_{i}(t)$ with respect to
time.

In this paper, we study the transient behaviors and properties
when the signal field (related to $\Omega_{24}$) is present,
particularly shortly after it's switched on. Thus, we should first
give the initial conditions for the probability amplitudes when
the signal light is absent. It is assumed that the present
four-level N-type EIT system can be reduced to a standard
three-level $\Lambda$-type EIT system ({\it i.e.},
$\Omega_{24}=0$) before the signal light is switched on. Note that
the coupling laser is strong, monochromatic and present for all
time. Therefore, when $|\Omega_{\rm p}|\ll|\Omega_{\rm c}|$
(always true in a standard three-level EIT system and the present
four-level EIT system), due to the quantum interference effect,
this three-level system will be nearly transparent (no absorption)
to the probe beam even if the probe beam is resonant with the
level transition $|1\rangle \rightarrow |3\rangle$. In other
words, if the intensity of the probe laser beam is sufficiently
weak, virtually all the atoms remain in the ground state, {\it
i.e.}, the atomic population in level $|1\rangle$ is
${a}_{1}\simeq 1$ (and ${a}_{3}$ is nearly vanishing).
Furthermore, $\dot{a}_{1}=({i}/{2})\Omega_{\rm p}a_{3}$ is
negligibly small since both $\Omega_{\rm p} $ and $a_{3}$ are
small compared with $\Omega_{\rm c} $ and $a_{1}$, respectively.
Thus, it is reasonable to assume that ${a}_{1}\simeq 1$ still
holds when considering the transient behaviors in the four-level
atomic medium induced by the quantum interference effect. In the
rest of the paper  we set ${a}_{1}= 1$. The equations for the
probability amplitudes ${a}_{2}$, ${a}_{3}$ and ${a}_{4}$ in
(\ref{eqamplitude}) can be rewritten in the following matrix form

\begin{equation}
\frac{\partial}{\partial t}\left( {\begin{array}{*{20}c}
   {a_{2}(t)}  \\
   {a_{3}(t)}  \\
    {a_{4}(t)}                   \\
\end{array}} \right) = \left( {\begin{array}{*{20}c}
   {-\frac{\gamma_{21}}{2}} & {\frac{i}{2}\Omega_{\rm c}} & {\frac{i}{2}\Omega_{\rm 24}}  \\
    {\frac{i}{2}\Omega^{\ast}_{\rm c}} & {-\frac{\Gamma_{3}}{2}} & {0} \\
    {\frac{i}{2}\Omega^{\ast}_{\rm 24}} & {0} & {-\frac{\Gamma_{4}}{2}}  \\
\end{array}} \right)\left( {\begin{array}{*{20}c}
   {a_{2}(t)}  \\
   {a_{3}(t)}  \\
    {a_{4}(t)}  \\
\end{array}} \right)+\left( {\begin{array}{*{20}c}
   {0}  \\
   {\frac{i}{2}\Omega_{\rm p}^{\ast}}  \\
    {0}  \\
\end{array}} \right).
\label{eqevolution}
\end{equation}
In order to solve Eq. (\ref{eqevolution}), one should first obtain
the eigenvalues $\lambda$ of the $3\times 3$ coefficient matrix in
Eq. (\ref{eqevolution}). These eigenvalues should satisfy
\begin{equation}
 {\rm det}\left( {\begin{array}{*{20}c}
   {-\frac{\gamma_{21}}{2}}-\lambda & {\frac{i}{2}\Omega_{\rm c}} & {\frac{i}{2}\Omega_{\rm 24}}  \\
    {\frac{i}{2}\Omega^{\ast}_{\rm c}} & {-\frac{\Gamma_{3}}{2}}-\lambda & {0} \\
    {\frac{i}{2}\Omega^{\ast}_{\rm 24}} & {0} & {-\frac{\Gamma_{4}}{2}}-\lambda  \\
\end{array}} \right)=0,           \label{determinant1}
\end{equation}
where ${\rm det}$ denotes the determinant of the matrix. Eq.
(\ref{determinant1}) gives the following cubic equation
\begin{equation}
\lambda^{3}+3b\lambda^{2}+3c\lambda+d=0,             \label{cubic}
\end{equation}
where
\begin{equation}
\left\{
\begin{array}{ll}
&   b=\frac{\gamma_{21}+\Gamma_{3}+\Gamma_{4}}{6},               \\
& c=\frac{\gamma_{21}\Gamma_{3}+\Gamma_{4}\left(\gamma_{21}+\Gamma_{3}\right)+\Omega^{\ast}_{\rm c}\Omega_{\rm c}+\Omega^{\ast}_{\rm 24}\Omega_{\rm 24}}{12},    \\
& d=\frac{\Gamma_{4}\left(\gamma_{21}\Gamma_{3}+\Omega^{\ast}_{\rm
c}\Omega_{\rm c}\right)+\Gamma_{3}\Omega^{\ast}_{\rm
24}\Omega_{\rm 24}}{8}.
\end{array}
\right.
\end{equation}
The three roots, $\lambda_{n}$, of the cubic equation
(\ref{cubic}) are
\begin{equation}
\left\{
\begin{array}{ll}
\lambda_{1} = u+v-b,  \\
\lambda_{2} = uw+vw^{2}-b,  \\
\lambda_{3} = uw^{2}+vw-b,
\end{array}
\right.
\end{equation}
where $u=[(-q+\sqrt{\Delta})/2]^{-1/3}$, $v=-p/u$,
$w=(-1+i\sqrt{3})/2$, $\Delta=4p^{3}+q^{2}$, $p=c-b^{2}$ and
$q=d-3bc+2b^{3}$.

We need to solve first the homogeneous equation corresponding to
Eq. (\ref{eqevolution}). Insertion of
$a^{(n)}_{2}(t)=a^{(n)}_{2}(0)\exp(\lambda_{n}t)$,
$a^{(n)}_{3}(t)=a^{(n)}_{3}(0)\exp(\lambda_{n}t)$,
$a^{(n)}_{4}(t)=a^{(n)}_{4}(0)\exp(\lambda_{n}t)$ ($n=1,2,3$) into
the homogeneous counterpart of Eq. (\ref{eqevolution}) yields the
following relations
\begin{equation}
a^{(n)}_{3}(0)=\frac{i\Omega^{\ast}_{\rm
c}}{\Gamma_{3}+2\lambda_{n}}a^{(n)}_{2}(0), \quad
a^{(n)}_{4}(0)=\frac{i\Omega^{\ast}_{\rm
24}}{\Gamma_{4}+2\lambda_{n}}a^{(n)}_{2}(0)   \quad  (n=1,2,3)
\end{equation}
between the coefficients of the general solutions of the
homogeneous equation. The above equations imply that both
$a^{(n)}_{3}(0)$ and $a^{(n)}_{4}(0)$ ($n=1,2,3$) can be expressed
in terms of $a^{(n)}_{2}(0)$ ($n=1,2,3$). Therefore, the general
solution of Eq. (\ref{eqevolution}) takes the form
\begin{equation}
\left( {\begin{array}{*{20}c}
   {a_{2}(t)}  \\
   {a_{3}(t)}  \\
    {a_{4}(t)}  \\
\end{array}} \right)=\sum_{n=1,2,3}a^{(n)}_{2}(0)\left( {\begin{array}{*{20}c}
   {1}  \\
   {\frac{i\Omega^{\ast}_{\rm
c}}{\Gamma_{3}+2\lambda_{n}}}  \\
    {\frac{i\Omega^{\ast}_{\rm
24}}{\Gamma_{4}+2\lambda_{n}}}  \\
\end{array}} \right)\exp(\lambda_{n}t)+{\bf a}_{({\rm s})},
\label{eqeq}
\end{equation}
where the column vector ${\bf a}_{({\rm s})}$ is a particular
solution (steady solution) of Eq. (\ref{eqevolution}), which can
be easily obtained by setting $\dot{a}_{2}=0$, $\dot{a}_{3}=0$ and
$\dot{a}_{4}=0$ in Eq. (\ref{eqamplitude}). The result is given as
follows
\begin{equation}
{\bf a}_{({\rm s})}=\frac{1}{{\Omega^{\ast}_{\rm c}\Omega_{\rm
c}{\Gamma_{4}}+\gamma_{21}\Gamma_{3}\Gamma_{4}+{\Omega^{\ast}_{\rm
24}\Omega_{\rm 24}\Gamma_{3}}}}\left( {\begin{array}{*{20}c}
   {{-\Omega^{\ast}_{\rm
p}\Omega_{\rm c}{\Gamma_{4}}}}  \\
   {{i\Omega^{\ast}_{\rm
p}\left(\Omega^{\ast}_{\rm 24}\Omega_{\rm
24}+\gamma_{21}\Gamma_{4}\right)}}  \\
    {{-i\Omega^{\ast}_{\rm
p}\Omega_{\rm c}\Omega^{\ast}_{\rm 24}}}  \\
\end{array}} \right).    \label{static}
\end{equation}

The three parameters $a^{(n)}_{2}(0)$ ($n=1,2,3$) in the general
solution (\ref{eqeq}) should be determined from the initial
condition for the probability amplitudes $a_{2}(0)$, $a_{3}(0)$
and $a_{4}(0)$ of levels $|2\rangle, |3\rangle $ and $|4\rangle$
through the following matrix equation
\begin{equation}
\left( {\begin{array}{*{20}c}
   {a_{2}(0)-a_{2({\rm s})}}  \\
   {a_{3}(0)-a_{3({\rm s})}}  \\
    {a_{4}(0)-a_{4({\rm s})}}  \\
\end{array}} \right)=\left(\begin{array}{cccc}
{1} & {1} & {1} \\
{\frac{i\Omega^{\ast}_{\rm c}}{\Gamma_{3}+2\lambda_{1}}} &
{\frac{i\Omega^{\ast}_{\rm c}}{\Gamma_{3}+2\lambda_{2}}} &
{\frac{i\Omega^{\ast}_{\rm
c}}{\Gamma_{3}+2\lambda_{3}}} \\
 {\frac{i\Omega^{\ast}_{\rm
24}}{\Gamma_{4}+2\lambda_{1}}} & {\frac{i\Omega^{\ast}_{\rm
24}}{\Gamma_{4}+2\lambda_{2}}} & {\frac{i\Omega^{\ast}_{\rm
24}}{\Gamma_{4}+2\lambda_{3}}}  \\
 \end{array}
 \right)\left( {\begin{array}{*{20}c}
   {a_{2}^{(1)}(0)}  \\
   {a_{2}^{(2)}(0)}  \\
    {a_{2}^{(3)}(0)}  \\
\end{array}} \right).      \label{eq211}
\end{equation}

First we consider the initial condition for the probability
amplitudes before the signal field is switched on. In such a
$\Lambda$-type EIT system, the equation of motion for probability
amplitudes ${a}_{2}$ and ${a}_{3}$ may be rewritten as
\begin{equation}
\dot{a}_{2}=\frac{i}{2}\Omega_{\rm
c}a_{3}-\frac{\gamma_{21}}{2}a_{2},  \quad
\dot{a}_{3}=\frac{i}{2}\left(\Omega^{\ast}_{\rm
c}a_{2}+\Omega^{\ast}_{\rm
p}a_{1}\right)-\frac{\Gamma_{3}}{2}a_{3}.
\end{equation}
The steady-state solution (for which $\dot{a}_{2}=\dot{a}_{3}=0$)
of the above set of equations is
\begin{equation}
a_{2\rm (s)}=-\frac{\Omega^{\ast}_{\rm p}\Omega_{\rm
c}}{\Omega^{\ast}_{\rm c}\Omega_{\rm
c}+\gamma_{21}\Gamma_{3}}\simeq -\frac{\Omega^{\ast}_{\rm
p}}{\Omega^{\ast}_{\rm c}},      \quad    a_{3\rm
(s)}=\frac{i\gamma_{21}\Omega^{\ast}_{\rm p}}{\Omega^{\ast}_{\rm
c}\Omega_{\rm c}+\gamma_{21}\Gamma_{3}}.   \label{obtained}
\end{equation}

Since the dephasing rate $\gamma_{21}$ is very small (less than
$\Gamma_{3,4}$ and $|\Omega_{\rm c}|$ by two orders of magnitude),
the steady value for $a_{3}$ is negligibly small. Thus, we set
zero initial value for $a_{3}$. From the dark-state condition for
a three-level $\Lambda$-type EIT system, one obtains the initial
condition for $a_{2}$: $a_{2}(0)=-\Omega^{\ast}_{\rm
p}/\Omega^{\ast}_{\rm c}$. Obviously, the initial condition for
$a_{4}$ is $a_{4}(0)=0$. These initial conditions will be used in
the following sections to determine the coefficients
$a_{2}^{(n)}(0)$. Thus, based on Eqs. (\ref{eqeq})-(\ref{eq211}),
one can obtain the time-dependent expression for the
susceptibility at the probe frequency\cite{Li}
\begin{equation}
\chi_{31}(t)=\frac{2N|\mu_{31}|^{2}}{\epsilon_{0}\hbar\Omega^{\ast}_{\rm
p}}\rho_{31}(t),    \label{susceptib}
\end{equation}
where $N$ and $\mu_{31}$ denote the atomic number per volume and
the atomic transition dipole between levels $|1\rangle$ and
$|3\rangle$, respectively, and the density matrix element
$\rho_{31}(t)=a_{3}(t)a^{\ast}_{1}(t)\simeq a_{3}(t)$.

In the next two sections, we will consider two special cases (for
which simpler formulae and clearer physical interpretations can be
obtained), namely, case when $|\Omega_{24}|\ll |\Omega_{\rm c}|$
({\it i.e.}, with a weak signal field) and case when
$|\Omega_{24}|\gg |\Omega_{\rm c}|$ ({\it i.e.}, with a strong
signal field). Note that in these two cases the conditions
($|\Omega_{\rm p}|, \gamma_{21}, \Gamma_{3,4}\ll |\Omega_{\rm
c}|$) are always satisfied.

\section{Influence of a weak signal field on the susceptibility for the probe light}

In this section we study the transient behaviors and properties
when $|\Omega_{24}|\ll |\Omega_{\rm c}|$. As discussed before, the
nonradiative transition rate $\gamma_{21}$ between $|1\rangle$ and
$|2\rangle$ is usually negligibly small. Thus, we can assume that
$\gamma_{21}\ll\Gamma_{3}$ (in general, $\gamma_{21}$ is two
orders of magnitude less than $\Gamma_{3}$). With this condition,
the three eigenvalues $\lambda_{n}$ of the coefficient matrix of
Eq. (\ref{eqevolution}) become
\begin{equation}
\lambda_{1}=-\frac{\Gamma_{4}}{2},  \quad
\lambda_{2}=\frac{-\Gamma_{3}+i\Omega}{4},   \quad
\lambda_{3}=\frac{-\Gamma_{3}-i\Omega}{4},
\end{equation}
where $\Omega=\sqrt{4\Omega^{\ast}_{\rm c}\Omega_{\rm
c}-\Gamma_{3}^{2}}$.

Since $|\Omega_{24}|\ll |\Omega_{\rm c}|$, the matrix elements
${i\Omega^{\ast}_{\rm 24}}/\left({\Gamma_{4}+2\lambda_{2}}\right)$
and ${i\Omega^{\ast}_{\rm
24}}/\left({\Gamma_{4}+2\lambda_{3}}\right)$ in Eq. (\ref{eq211})
is very small (in the order of $|\Omega^{\ast}_{\rm
24}/\Omega_{\rm c}|$). Thus, these two matrix elements can be set
to zero. Eq. (\ref{eq211}) can then be rewritten as
\begin{equation}
\left\{
\begin{array}{ll}
&  -\frac{\Omega^{\ast}_{\rm p}}{\Omega^{\ast}_{\rm
c}}-a_{2(\rm s)}=a_{2}^{(1)}(0)+a_{2}^{(2)}(0)+a_{2}^{(3)}(0),              \\
&  -a_{3(\rm s)}=\frac{i\Omega^{\ast}_{\rm
c}}{\Gamma_{3}+2\lambda_{1}}a_{2}^{(1)}(0)+\frac{i\Omega^{\ast}_{\rm
c}}{\Gamma_{3}+2\lambda_{2}}a_{2}^{(2)}(0)+\frac{i\Omega^{\ast}_{\rm
c}}{\Gamma_{3}+2\lambda_{3}}a_{2}^{(3)}(0),             \\
 &  -a_{4(\rm s)}\simeq \frac{i\Omega^{\ast}_{\rm
24}}{\Gamma_{4}+2\lambda_{1}}a_{2}^{(1)}(0),
\end{array}
\right.      \label{determine.coefficients}
\end{equation}
which can be used to determine $a^{(n)}_{2}(0)$ ($n=1,2,3$) in the
solution (\ref{eqeq}). According to the steady-state solution
(\ref{static}), $a_{2(\rm s)}$ is approximately
$-{\Omega^{\ast}_{\rm p}}/{\Omega^{\ast}_{\rm c}}$ when
$\gamma_{21}\ll\Gamma_{3}$ and $|\Omega_{24}|\ll |\Omega_{c}|$.
Using this result and relations $\lambda_{1}=-\Gamma_{4}/2$,
$\Gamma_{3}+2\lambda_{3}=-2\lambda_{2}$ and
$\Gamma_{3}+2\lambda_{2}=-2\lambda_{3}$, one obtains
\begin{equation}
\left\{
\begin{array}{ll}
&  a_{2}^{(1)}(0)\simeq
\frac{i\left(\Gamma_{4}+2\lambda_{1}\right)a_{4(\rm
s)}}{\Omega^{\ast}_{\rm 24}}=0,
 \\
 &    a_{2}^{(3)}(0)=\frac{4a_{3(\rm
s)}\lambda_{2}\lambda_{3}+2i\Omega^{\ast}_{\rm
c}\lambda_{2}\left(\frac{\Omega^{\ast}_{\rm p}}{\Omega^{\ast}_{\rm
c}}+a_{2(\rm s)}\right)}{\Omega^{\ast}_{\rm
c}\Omega}=\frac{4a_{3(\rm
s)}\lambda_{2}\lambda_{3}}{\Omega^{\ast}_{\rm c}\Omega},             \\
 &   a_{2}^{(2)}(0)=-\frac{\Omega^{\ast}_{\rm
p}}{\Omega^{\ast}_{\rm c}}-a_{2(\rm
s)}-a_{2}^{(3)}(0)=-a_{2}^{(3)}(0).
\end{array}
\right.            \label{arriveat}
\end{equation}
It thus follows from  Eq. (\ref{eqeq}) that the time-dependent
expression for the probability amplitude of level $|3\rangle$ is
\begin{eqnarray}
a_{3}(t)&=&-a_{2}^{(3)}(0)i\Omega^{\ast}_{\rm c}\left[\frac{\exp(\lambda_{2}t)}{-2\lambda_{3}}-\frac{\exp(\lambda_{3}t)}{-2\lambda_{2}}\right]+a_{3({\rm s})}              \nonumber \\
&=& a_{3({\rm
s})}\left\{-\exp\left(-\frac{\Gamma_{3}}{4}t\right)\left[\cos\left(\frac{\Omega}{4}t\right)
-\frac{\Gamma_{3}}{\Omega}\sin\left(\frac{\Omega}{4}t\right)\right]+1\right\}.
\label{a3}
\end{eqnarray}
In a similar way, we obtain the following transient probability
amplitude of ground state $|2\rangle$
\begin{equation}
a_{2}(t)=-2ia_{3(\rm s)}\frac{\Omega_{\rm
c}}{\Omega}\exp\left(-\frac{\Gamma_{3}}{4}t\right)\sin\left(\frac{\Omega}{4}t\right)+a_{2(\rm
s)}.     \label{a2}
\end{equation}

According to Eq. (\ref{eqamplitude}), the transient probability
amplitude of excited level $|4\rangle$ in the presence of the
signal field $\Omega_{24}$ satisfies
$\dot{a}_{4}=({i}/{2})\Omega^{\ast}_{24}a_{2}-({\Gamma_{4}}/{2})a_{4}$.
Substituting expression (\ref{a2}) into this equation, we obtain
\begin{eqnarray}
{a}_{4}(t) &=&
\exp\left(-\frac{\Gamma_{4}}{2}t\right)\left\{a_{3(\rm
s)}\frac{\Omega^{\ast}_{24}\Omega_{\rm
c}}{\Omega}\int^{t}_{0}\exp\left[\left(\frac{\Gamma_{4}}{2}
-\frac{\Gamma_{3}}{4}\right)t'\right]\sin\left(\frac{\Omega}{4}t'\right){\rm
d}t'+a_{2(\rm
s)}\frac{i\Omega^{\ast}_{24}}{\Gamma_{4}}\left[\exp\left(\frac{\Gamma_{4}}{2}t\right)-1\right]+{\mathcal
C}\right\}      \nonumber \\
&=& a_{3(\rm s)}\frac{\Omega^{\ast}_{24}\Omega_{\rm c}}{\Omega
\left[\left(\frac{\Gamma_{4}}{2}-\frac{\Gamma_{3}}{4}\right)^{2}+\left(\frac{\Omega}{4}\right)^{2}\right]}\left\{
\exp\left(-\frac{\Gamma_{3}}{4}t\right)\left[\left(\frac{\Gamma_{4}}{2}-\frac{\Gamma_{3}}{4}\right)\sin\left(\frac{\Omega}{4}t\right)-\frac{\Omega}{4}\cos\left(\frac{\Omega}{4}t\right)
\right]+\frac{\Omega}{4}\exp\left(-\frac{\Gamma_{4}}{2}t\right)
\right\}
 \nonumber \\
& &  +a_{2(\rm
s)}\left(\frac{i\Omega^{\ast}_{24}}{\Gamma_{4}}\right)\left[1-\exp
\left(-\frac{\Gamma_{4}}{2}t\right)\right],   \label{a4}
\end{eqnarray}
where the integral constant ${\mathcal C}$ should be zero due to
the initial condition ${a}_{4}(0)=0$. It is readily verified that
the steady value (when $t \rightarrow \infty$) of ${a}_{4}$ is
$({i\Omega^{\ast}_{24}}/{\Gamma_{4}})a_{2(\rm s)}$, which agrees
with the steady value (\ref{static}). Thus we obtain the explicit
expressions for the probability amplitudes of levels $|2\rangle$,
$|3\rangle$ and $|4\rangle$ in the transient evolution process of
the four-level N-type EIT media. Such a transient evolution
process commences as the signal laser is switched on.

Next we discuss the influence of the signal field on the induced
polarizability of the EIT medium due to the
$|1\rangle$-$|3\rangle$ transition and the nonlinear absorption
for the probe light. Insertion of Eq. (\ref{a3}) into Eq.
(\ref{susceptib}) yields
\begin{equation}
\chi_{31}(t)=\frac{2N|\mu_{31}|^{2}}{\epsilon_{0}\hbar\Omega^{\ast}_{\rm
p}}a_{3({\rm
s})}\left\{-\exp\left(-\frac{\Gamma_{3}}{4}t\right)\left[\cos\left(\frac{\Omega}{4}t\right)
-\frac{\Gamma_{3}}{\Omega}\sin\left(\frac{\Omega}{4}t\right)\right]+1\right\},
\label{nonlinear}
\end{equation}
where the steady probability amplitude $a_{3({\rm s})}$ is given
by (\ref{static}) (the second element).

As an illustrative example, we choose the following values
(typical for transitions in hyperfine-split Na D lines
\cite{Cowan}): $\Gamma_{3}=1.2\times 10^{8}$ s$^{-1}$,
$\Gamma_{4}=2.5\times 10^{8}$ s$^{-1}$ and $\gamma_{21}\simeq
3\times 10^{6}$ s$^{-1}$. The susceptibility $\chi_{31}(t)$ at the
probe frequency is purely imaginary (see the above equation; note
that $ a_{3({\rm s})}$ is purely imaginary) and the transient
behavior of ${\rm Im}\{\chi_{31}\}$ is shown in Fig. 2 for various
values of $\Omega_{\rm c}$ ($\Omega_{24} = 0.1 \Omega_{\rm c}$).
At $t=0$, the susceptibility is zero since before the signal field
$\Omega_{\rm 24}$ is switched on the four-level N-type atomic
ensemble can be reduced to a three-level $\Lambda$-type EIT
system. However, once the signal field is switched on, the
susceptibility $\chi_{31}(t)$ (and hence the nonlinear absorption)
for the probe beam in such a four-level N-type system is
oscillating (with damped oscillating amplitude) but will finally
reach the steady-state value. If the Rabi frequency $\Omega_{\rm
c}$ of the coupling beam becomes greater, the susceptibility curve
for the probe laser will oscillate more significantly (cf. the
solid curve of Fig. 2). When $\Omega_{\rm c}$ decreases to
$2\Gamma_{3}$, one has $\Omega =0$ and thus the oscillation in the
susceptibility curve vanishes (cf. the dotted curve of Fig. 2).

The steady value of the susceptibility for the probe beam is
\begin{equation}
\chi_{31}(\infty)=\frac{i2N|\mu_{31}|^{2}\left(\Omega^{\ast}_{\rm
24}\Omega_{\rm
24}+\gamma_{21}\Gamma_{4}\right)}{\epsilon_{0}\hbar\left({\Omega^{\ast}_{\rm
c}\Omega_{\rm
c}{\Gamma_{4}}+\gamma_{21}\Gamma_{3}\Gamma_{4}+{\Omega^{\ast}_{\rm
24}\Omega_{\rm 24}\Gamma_{3}}}\right)}.    \label{steady}
\end{equation}
The four-level EIT medium is absorptive for the probe beam since
the susceptibility at the probe frequency is purely imaginary (cf.
the above two equations). From expression (\ref{steady}) one sees
that if the dephasing rate $\gamma_{21}\rightarrow 0$ the linear
absorption for the probe beam vanishes. However, due to the
presence of the signal field, there appears a nonlinear absorption
which is represented by
\begin{equation}
{\rm
Im}\{\chi_{31}^{(3)}\}(\infty)=\frac{2N|\mu_{31}|^{2}\Omega^{\ast}_{\rm
24}\Omega_{\rm 24}}{\epsilon_{0}\hbar\left({\Omega^{\ast}_{\rm
c}\Omega_{\rm c}{\Gamma_{4}}+{\Omega^{\ast}_{\rm 24}\Omega_{\rm
24}\Gamma_{3}}}\right)}.   \label{Im}
\end{equation}
Since $\gamma_{21}\ll\Gamma_{3}$ and $|\Omega_{24}|\ll
|\Omega_{c}|$, expression (\ref{Im}) can be reduced to
\begin{equation}
{\rm Im}\{\chi_{31}^{(3)}\}(\infty)\simeq
\frac{2N|\mu_{31}|^{2}}{\epsilon_{0}\hbar\Omega^{\ast}_{\rm
p}}\left(\frac{\Omega^{\ast}_{\rm
p}}{\Gamma_{4}}\right)\left|\frac{\Omega_{\rm 24}}{\Omega_{\rm
c}}\right|^{2}. \label{reduced}
\end{equation}
In a two-level system, however, the imaginary part of the linear
susceptibility $\chi^{(1)}$ can be written in the following form
\begin{equation}
{\rm Im}\{\chi^{(1)}\}(\infty)=
\frac{2N|\mu|^{2}}{\epsilon_{0}\hbar\Omega^{\ast}}\left(\frac{\Omega^{\ast}}{\Gamma}\right),
\label{twolevel}
\end{equation}
where $\Omega$ denotes the resonant laser field coupled to the
two-level system, and $\mu$ and $\Gamma$ the dipole matrix element
and the decay rate of such an atomic system, respectively. From
Eqs. (\ref{reduced}) and (\ref{twolevel}), one can see that the
nonlinear absorption coefficient for the probe light in a
four-level system has a similar form except an additional factor
$|\Omega_{\rm 24}/\Omega_{\rm c}|^{2}$ as compared with the linear
absorption coefficient for the resonant light in a two-level
system. Harris and Yan {\it et al.}\cite{Yamamoto,Yan} have
pointed out that due to the EIT cancellation of single-photon
absorption the nonlinear photon absorption is dramatically
enhanced and the observed nonlinear absorption amplitude  may
become comparable to that of the single-photon absorption in a
two-level system. This is consistent with our theoretical results.
Since the linear polarizability in the four-level EIT atomic
medium is eliminated by the quantum interference (in the
meanwhile, the linear absorption associated with the dephasing
rate $\gamma_{21}$ has been ignored due to the smallness of
$\gamma_{21}$), the above nonlinear absorption effect may dominate
the optical behaviors and properties of such a multilevel EIT
medium. Moreover, the signal field will greatly enhance the
nonlinear absorption. Thus, such a system may function as a
two-photon absorptive optical switch by turning on and off the
signal field \cite{Yamamoto}.

\section{Influence of a strong signal field on the susceptibility for the probe light}

In this section, we will consider the transient evolutional
behavior of the four-level EIT medium in another case when
$|\Omega_{24}|\gg |\Omega_{\rm c}|$ ({\it i.e.}, with a strong
signal field). For this case, the eigenvalues of the coefficient
matrix in Eq. (\ref{eqevolution}) are \begin{equation}
\lambda_{1}=-\frac{\Gamma_{3}}{2},   \quad
\lambda_{2}=\frac{-\Gamma_{4}+i\Omega'}{4}, \quad
\lambda_{3}=\frac{-\Gamma_{4}-i\Omega'}{4}, \label{eigenvalue2}
\end{equation}
where
$\Omega'=\sqrt{4\Omega^{\ast}_{24}\Omega_{24}-\Gamma_{4}^{2}}$.
Substituting the initial conditions ({\it i.e.}, $a_{1}(0)=1$,
$a_{2}(0)=-\Omega^{\ast}_{\rm p}/\Omega^{\ast}_{\rm c}$,
$a_{3}(0)=0$ and $a_{4}(0)=0$) for the probability amplitudes
 into
Eq. (\ref{eq211}), we obtain the following equations
\begin{equation} \left\{
\begin{array}{ll}
 &      -\frac{\Omega^{\ast}_{\rm p}}{\Omega^{\ast}_{\rm
c}}-a_{2(\rm s)}=a_{2}^{(1)}(0)+a_{2}^{(2)}(0)+a_{2}^{(3)}(0),             \\
 &     -a_{3(\rm s)}\simeq \frac{i\Omega_{\rm c}^{\ast}}{\Gamma_{3}+2\lambda_{1}}a^{(1)}_{2}(0),           \\

 &    -a_{4(\rm s)}=a_{2}^{(1)}(0)\frac{i\Omega^{\ast}_{\rm
24}}{\Gamma_{4}+2\lambda_{1}}+a_{2}^{(2)}(0)\frac{i\Omega^{\ast}_{\rm
24}}{\Gamma_{4}+2\lambda_{2}}+a_{2}^{(3)}(0)\frac{i\Omega^{\ast}_{24}}{\Gamma_{4}+2\lambda_{3}}
\end{array}
\right.     \label{coefficientequation}
\end{equation}
for the coefficients $a_{2}^{(n)}(0)$ ($n=1,2,3$). Insertion of
the eigenvalues (\ref{eigenvalue2}) into Eqs.
(\ref{coefficientequation}) gives the explicit expressions for the
three coefficients $a_{2}^{(n)}$ ($n=1,2,3$), {\it i.e.},
\begin{equation}
\left\{
\begin{array}{ll}
 &   a_{2}^{(1)}(0)\simeq
\frac{i\left(\Gamma_{3}+2\lambda_{1}\right)a_{3(\rm
s)}}{\Omega_{\rm c}^{\ast}}=0, \\
& a_{2}^{(2)}(0)=\frac{-2a_{4(\rm s)}\Omega_{24}-2a_{2(\rm
s)}\Omega'+i\left(\Gamma_{4}-i\Omega'\right)\left(\frac{\Omega^{\ast}_{\rm
p}}{\Omega^{\ast}_{\rm c}}+a_{2(\rm
s)}\right)}{2\Omega'}-\frac{\Omega^{\ast}_{\rm
p}}{\Omega^{\ast}_{\rm c}},  \\
 &  a_{2}^{(3)}(0)=\frac{2a_{4(\rm
s)}\Omega_{24}-i\left(\Gamma_{4}-i\Omega'\right)\left(\frac{\Omega^{\ast}_{\rm
p}}{\Omega^{\ast}_{\rm c}}+a_{2(\rm s)}\right)}{2\Omega'}.
\end{array}
\right.       \label{eq43}
\end{equation}
Substitution of expressions (\ref{eq43}) into solutions
(\ref{eqeq}) will yield the explicit expressions for the
probability amplitudes of the atomic levels of the four-level EIT
media under the condition $|\Omega_{\rm c}|\ll |\Omega_{24}|$. For
example, the probability amplitude of level $|3\rangle$ is
\begin{eqnarray}
a_{3}(t)&=& -i\left[\frac{\Omega^{\ast}_{\rm
c}a_{2}^{(2)}(0)}{2\lambda_{2}+\Gamma_{3}}+\frac{\Omega^{\ast}_{\rm
c}a_{2}^{(3)}(0)}{2\lambda_{3}+\Gamma_{3}}+\left(\frac{\Omega^{\ast}_{\rm
c}a_{2(\rm s)}+\Omega^{\ast}_{\rm
p}}{\Gamma_{3}}\right)-\frac{\gamma_{21}\Omega^{\ast}_{\rm
p}}{\Omega^{\ast}_{\rm c}\Omega^{\ast}_{\rm
c}+\gamma_{21}\Gamma_{3}}\right]\exp\left(-\frac{\Gamma_{3}}{2}t\right)
 \nonumber \\
 & &  +i\Omega^{\ast}_{\rm c}\left[\frac{a_{2}^{(2)}(0)}{2\lambda_{2}+\Gamma_{3}}\exp\left(\lambda_{2}t\right)
+\frac{a_{2}^{(3)}(0)}{2\lambda_{3}+\Gamma_{3}}\exp\left(\lambda_{3}t\right)\right]+i\left(\frac{\Omega^{\ast}_{\rm
c}a_{2(\rm s)}+\Omega^{\ast}_{\rm p}}{\Gamma_{3}}\right).
\label{a3expression}
\end{eqnarray}
Then from the explicit expression (\ref{a3expression}) for $a_{3}$
and Eq. (\ref{susceptib}) we can obtain the transient behavior of
the susceptibility $\chi_{31}(t)$ at the probe frequency. It is
shown that when $t \rightarrow \infty $ the steady value of the
susceptibility for the probe beam is
\begin{equation}
{\rm Im}\{\chi_{31}\}(\infty)\simeq {\rm
Im}\{\chi_{31}^{(1)}\}+{\rm Im}\{\chi_{31}^{(3)}\}+{\rm
Im}\{\chi_{31}^{(5)}\}+ ...=
\frac{2N|\mu_{31}|^{2}}{\epsilon_{0}\hbar\Omega^{\ast}_{\rm
p}}\left(\frac{\Omega^{\ast}_{\rm
p}}{\Gamma_{3}}\right)\left[1-\frac{\Omega^{\ast}_{\rm
c}\Omega_{\rm c}}{\Omega^{\ast}_{\rm 24}\Omega_{\rm
24}}\frac{\Gamma_{4}}{\Gamma_{3}}+\left(\frac{\Omega^{\ast}_{\rm
c}\Omega_{\rm c}}{\Omega^{\ast}_{\rm 24}\Omega_{\rm
24}}\frac{\Gamma_{4}}{\Gamma_{3}}\right)^{2}+...\right].
\label{eq45}
\end{equation}
It follows from the above formula that both the linear absorption
and nonlinear (various orders) absorption for the probe light
exist for the case of $|\Omega_{24}|\gg |\Omega_{\rm c}|$. This is
different from the absorptive behavior in the case of
$|\Omega_{24}|\ll |\Omega_{\rm c}|$ where the linear absorption
vanishes.

Fig. 3 shows the transient behavior of ${\rm Im}\{\chi_{31}\}$ for
various values of $\Omega_{\rm 24}$ (with $\Omega_{\rm
p}=0.1\Omega_{\rm c}$). The other parameters are the same as those
used in Fig. 2. As $\Omega_{\rm 24}$ becomes very large, ${\rm
Im}\{\chi_{31}\}$ approaches the same steady value
${2N|\mu_{31}|^{2}}/({\epsilon_{0}\hbar \Gamma_{3}})$ (cf.
Eq.(\ref{eq45})), which means the linear absorption for the probe
light. In other words, if the signal field is much stronger than
the coupling light, the nonlinear absorption for the probe light
may be greatly inhibited.

Comparing Eq. (\ref{coefficientequation}) with
(\ref{determine.coefficients}), one sees that the roles of
$a_{3}(t)$ and $a_{4}(t)$ in the case of $|\Omega_{24}|\gg
|\Omega_{\rm c}|$ are equivalent to those of $a_{4}(t)$ and
$a_{3}(t)$ in the case of $|\Omega_{24}|\ll |\Omega_{\rm c}|$,
respectively, namely, the transient behaviors and properties of
level $|3\rangle$ (or level $|4\rangle$) in the case of
$|\Omega_{24}|\gg |\Omega_{\rm c}|$ is similar to those of level
$|4\rangle$ (or level $|3\rangle$) in the case of
$|\Omega_{24}|\ll |\Omega_{\rm c}|$. Such a property may be called
{\it level reciprocity}, which is an interesting phenomenon
between the two excited states in the two cases in the four-level
N-type EIT system.

\section{Concluding remarks}

In the present paper, we have considered the transient evolutional
process of a four-level EIT medium when the signal field is
switched on.  Once the signal light is switched on, the atomic
population and the susceptibility (absorption) of the initial
state ({\it i.e.}, the three-level $\Lambda$-type EIT system)
oscillatorily approaches the steady values of a four-level EIT
system at a time scale of lifetime (several nanoseconds) of the
excited atomic levels. Both the linear and the nonlinear optical
properties of a multilevel atomic system can be modified by the
phase coherence and the quantum interference that utilizes EIT. We
have considered the transient process of establishing large
enhancement of the nonlinear polarizability in the four-level EIT
system: specifically, in the case of $|\Omega_{24}|\ll
|\Omega_{\rm c}|$, the linear absorption for the probe light
vanishes and the only retained absorption is of the third-order
nonlinearity; in the case of $|\Omega_{24}|\gg |\Omega_{\rm c}|$,
however, both the linear and the various-order nonlinear
absorptions exist. But once the signal field becomes stronger, the
various-order nonlinear absorptions will be inhibited and the
retained absorption is of the linearity only. In addition, we have
also found an interesting level-reciprocity relationship between
the two excited states in the two cases of $|\Omega_{24}|\gg
|\Omega_{\rm c}|$ and $|\Omega_{24}|\ll |\Omega_{\rm c}|$.
\\ \\
 \textbf{Acknowledgements}  This work is supported
by the National Natural Science Foundation of China under Project
Nos. $90101024$ and $60378037$.
\\ \\
\newpage

\section*{Figure Captions}

Fig. 1. Schematic diagram for a four-level N-type atomic system.
Levels $|2\rangle$ and $|3\rangle$ are coherently coupled by the
coupling beam with Rabi frequency $\Omega_{\rm c}$. The signal
field is switched on at $t=0$.
\\ \\

Fig. 2. The susceptibility ${\rm Im}\{\chi_{31}\}$ (in the unit of
$\frac{2N|\mu_{31}|^{2}}{\epsilon_{0}\hbar}$) as time increases
for the case $\Omega_{24}\ll \Omega_{\rm c}$. Here $\Omega_{24}=
0.1\Omega_{\rm c}$.
\\ \\

Fig. 3. The susceptibility ${\rm Im}\{\chi_{31}\}$ (in the unit of
$\frac{2N|\mu_{31}|^{2}}{\epsilon_{0}\hbar}$) as time increases
for the case $\Omega_{24}\gg \Omega_{\rm c}$. Here $\Omega_{\rm
p}= 0.1\Omega_{\rm c}$.

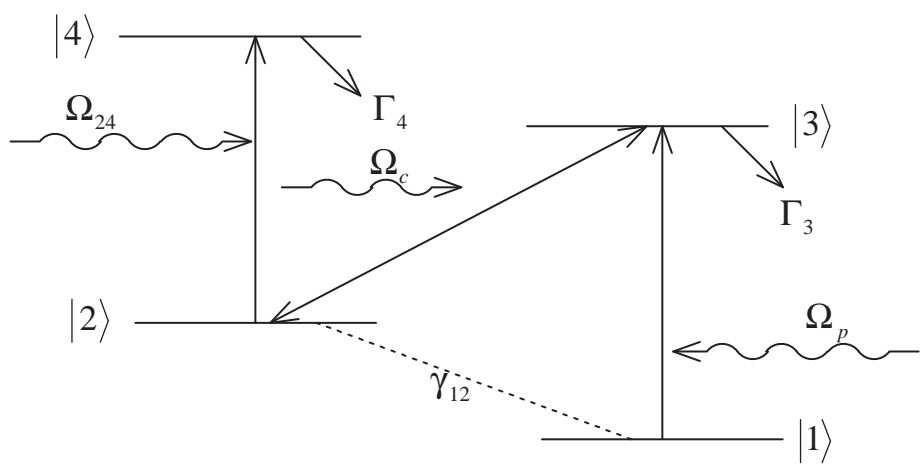
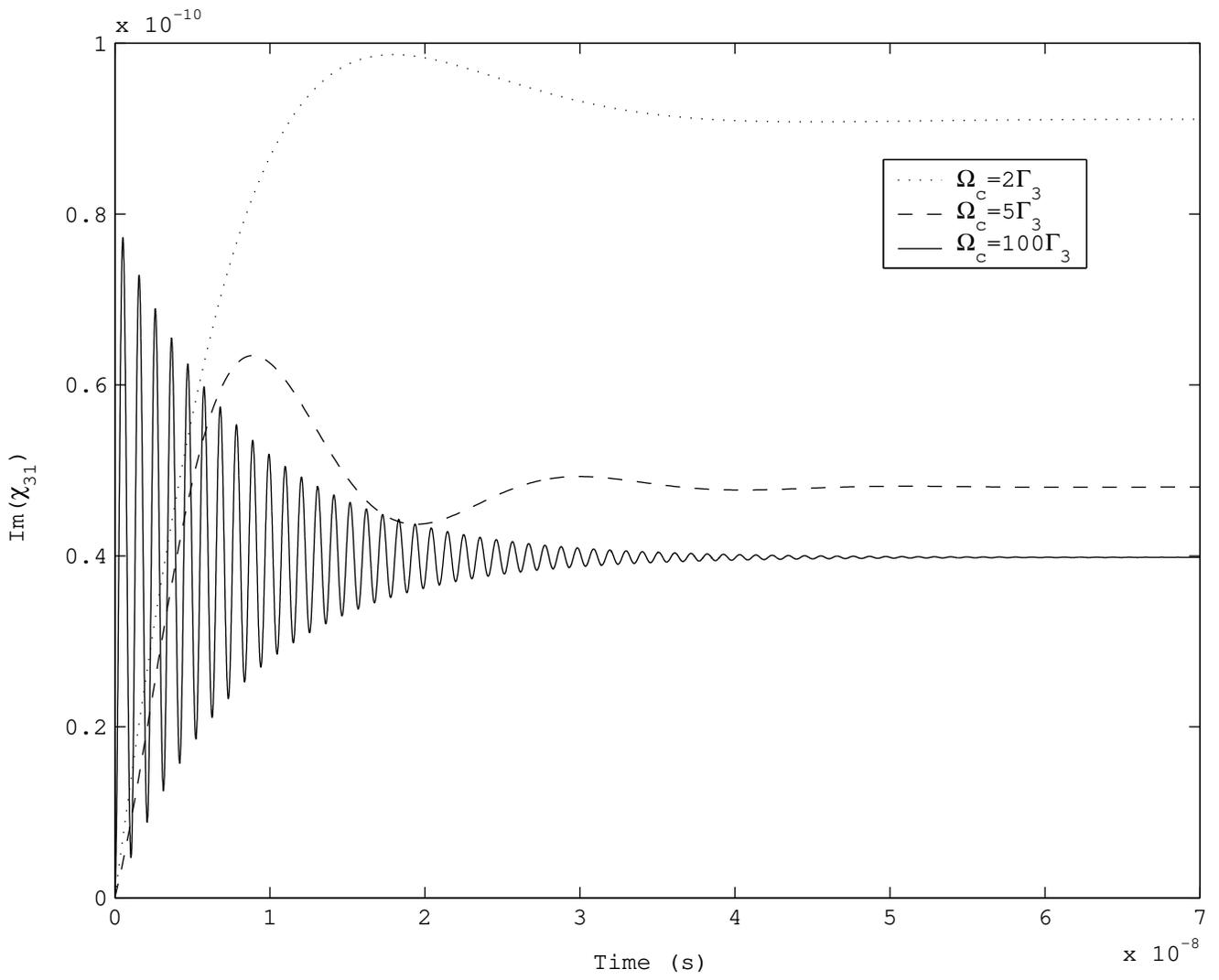
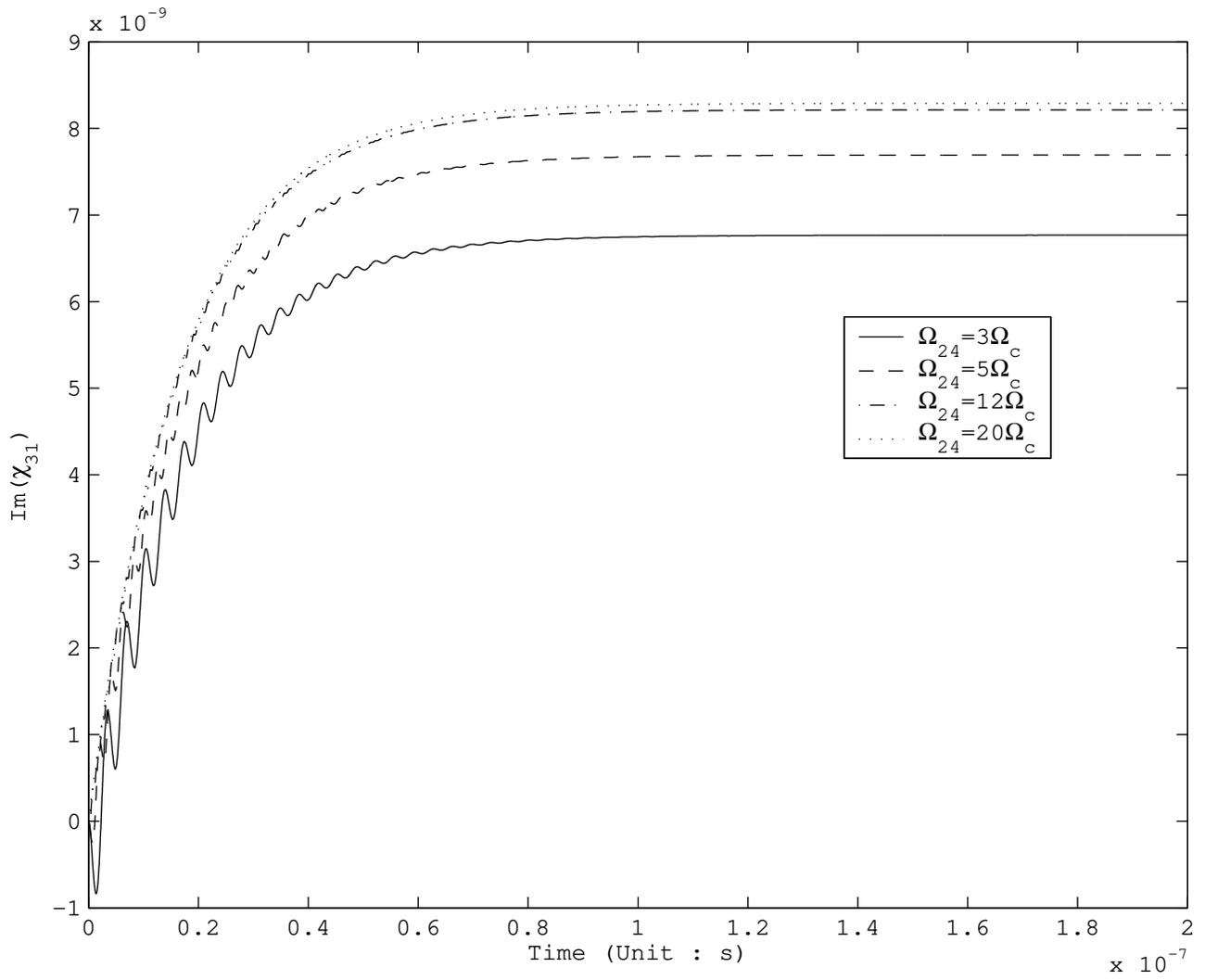
\end{document}